 \documentclass[12pt,draftclsnofoot,journal,onecolumn]{IEEEtran}
\usepackage{graphicx}
\usepackage{amsmath}
\usepackage{array}
\newcommand{\PreserveBackslash}[1]{\let\temp=\\#1\let\\=\temp}
\newcolumntype{C}[1]{>{\PreserveBackslash\centering}p{#1}}
\newcolumntype{R}[1]{>{\PreserveBackslash\raggedleft}p{#1}}
\newcolumntype{L}[1]{>{\PreserveBackslash\raggedright}p{#1}}
\usepackage[usenames]{color}
\usepackage{colortbl,booktabs}
\usepackage{tabularx}
\usepackage{multicol}
\usepackage{booktabs}
\usepackage[Symbol]{upgreek}
\usepackage{subfigure}
\usepackage{bm}
\usepackage{cite}
\usepackage{balance}
\usepackage[boxed]{algorithm2e}
\usepackage{verbatim}
\usepackage{threeparttable}
\usepackage{amsmath}
\usepackage{dcolumn}
\usepackage{multirow}
\usepackage{booktabs}
\usepackage{amsfonts}
\newcolumntype{d}[1]{D{.}{.}{#1}}

\begin{document}

\bibliographystyle{IEEEtran} 

\title{Hybrid Beamforming for Subarray-Level Movable Antenna Enhanced MU-MIMO Communications}

\author{Shanshan Zhang, Songjie Yang, Wenxuan Zhang, Youzhi Xiong, \emph{IEEE Member},
	and Siya Yao, \emph{IEEE Member}

\thanks{Shanshan Zhang and Siya Yao are with the College of Information and Electronic Engineering, Zhejiang Gongshang University, Hangzhou 310018, China (e-mail: 19012723396@163.com, yaosiya@zjgsu.edu.cn).

Songjie Yang is with the National Key Laboratory of Wireless Communications, University of Electronic Science and Technology of China, Chengdu 611731, China. (e-mail: yangsongjie@std.uestc.edu.cn)

Wenxuan Zhang is with the Department of Electrical and Computer Engineering, University of California, San Diego, La Jolla, CA 92093, USA. (e-mail: wez044@ucsd.edu)

Youzhi Xiong is with College of Physics and Electronic Engineering, Sichuan Normal University, Chengdu 610101, China. (e-mail: yzxiong@sicnu.edu.cn)}
}

\maketitle
\begin{abstract}
This study investigates subarray-level movable antenna (MA) architecture for multi-user MIMO (MU-MIMO) systems. Unlike conventional systems with fixed-position antennas (FPAs), the proposed scheme harnesses the additional positional degrees of freedom (DoFs) of movable subarrays to enhance spatial multiplexing capabilities for both multi-user and multi-stream communications. Our objective is to maximize the overall system spectral efficiency by jointly optimizing the hybrid beamforming design and the positions of all subarrays. To tackle this challenging non-convex optimization problem, we first adopt a block diagonalization (BD) based digital precoder to effectively eliminate multi-user interference. Subsequently, the joint optimization of the analog beamformer and the subarray positions is efficiently solved using a sequential interference cancellation (SIC)-based algorithm. Simulation results demonstrate that the proposed SIC-MA method significantly outperforms the benchmark SIC-FPA scheme where subarrays are fixed.

\end{abstract}


\section{Introduction}\label{S1}

The exploration of high-frequency bands, such as millimeter-wave and terahertz, is pivotal for beyond-5G systems to achieve ultra-high data rates. To combat the severe path loss in these bands, large-scale antenna arrays are essential for providing substantial beamforming gain \cite{busari2018survey,ning2026precoding,yang2025nearfieldestimation}. However, the hardware cost and power consumption of fully-digital architectures are prohibitive. Hybrid beamforming, employing a limited number of radio frequency (RF) chains, strikes a balance between performance and complexity \cite{ning2023thzbeamforming,yang2025nearfieldhybrid}. Particularly, the sub-connected structure, where each RF chain is connected to a sub-array of antennas, has been favored for its higher energy efficiency and easier implementation compared to the fully-connected counterpart \cite{yang2023multibeamtraining,zhang2025mahybrid}. Nonetheless, conventional hybrid beamforming schemes are built upon fixed-position antenna (FPA) arrays, whose performance is fundamentally limited by their inability to exploit the spatial degrees of freedom (DoFs) in the wireless channel \cite{zhu2024maopportunities}.

Recently, movable antenna (MA) technology has emerged as a paradigm-shifting solution to overcome this limitation \cite{zhu2026tutorial}. By enabling \textcolor{black}{flexible} position adjustments, MAs can dynamically reconfigure the electromagnetic field surrounding the transceiver, \textcolor{black}{creating} more favorable channel conditions. Primary works have established a field-response based channel model for MA systems and demonstrated their significant potential in improving received signal power \cite{zhu2024mamodeling}, enhancing MIMO capacity \cite{ma2024capacity,yang2025flexiblewmmse}, and facilitating multi-user communications \cite{zhu2024positionopt,yang2024flexibleprecoding}. \textcolor{black}{This has further spurred joint optimization of MA positions and other resources,} such as beamforming for physical-layer security \cite{hu2024securema} and sensing enhancement \cite{lyu2025maisac}.

However, the implementation of MA systems faces two primary challenges: high optimization complexity and stringent mutual coupling constraints. Existing literature has proposed a variety of algorithms to tackle the non-convex problem of MA position optimization. For instance, gradient ascent methods with backtracking line search are employed in \cite{cheng2024fluid}, while \cite{ma2024capacity} utilizes successive convex approximation (SCA) to handle the distance constraints. Alternatively, stochastic search algorithms like particle swarm optimization (PSO) have been applied \cite{xiao2024jointpositioning}, though they often suffer from \textcolor{black}{high overhead}. \textcolor{black}{Sparse optimization has also been used to reduce complexity} \cite{yang2024flexibleprecoding}.
Early MA studies presume a fully-digital architecture where each antenna element is connected to a dedicated RF chain and moved by an individual motor. This assumption leads to prohibitive cost and control complexity for systems with a large number of antennas, rendering such architectures impractical for widespread deployment. Therefore, array-level MAs are investigated in \cite{zhang2025mahybrid,ning2025architectures,yang2025flexiblearrays}. Particularly, \textcolor{black}{integrating MA into hybrid beamforming MU-MIMO is non-trivial due to the constant-modulus analog precoder constraints and inter-user interference. How to efficiently exploit subarray position flexibility to enhance spatial multiplexing under hybrid architectures remains open.}
\textcolor{black}{Motivated by this gap,} the hybrid beamforming architecture with MAs proposed in \cite{zhang2025mahybrid,xiang2026multibeam} introduces positional degrees of freedom by moving the subarrays. \textcolor{black}{However, the gain beyond conventional hybrid beamforming, especially for multi-stream MU-MIMO under sub-connected constraints, still needs careful validation.}

Inspired by the above, this work introduces subarray-level MAs into a multi-user MIMO (MU-MIMO) system. \textcolor{black}{We aim to maximize the spectral efficiency by jointly optimizing hybrid beamforming and subarray positions.} To tackle this challenging non-convex problem, we first adopt a block diagonalization (BD) precoding framework to eliminate multi-user interference. \textcolor{black}{We then optimize the analog beamformer and subarray positions via a sequential interference cancellation (SIC)-based approach.}

\textcolor{black}{
\noindent\textbf{Notations:} Scalars, vectors, and matrices are denoted by italic, bold lowercase, and bold uppercase letters, respectively. $(\cdot)^{T}$ and $(\cdot)^{H}$ denote the transpose and Hermitian transpose;  $\|\cdot\|_{F}$ denotes the  Frobenius norm; $|\cdot|$ represents the determinant of a matrix and the absolute value of a scalar; $\angle(\cdot)$ returns the element-wise phase; $\otimes$ denotes the Kronecker product; $\mathrm{blkdiag}(\cdot)$ constructs a block-diagonal matrix; $\mathbf{I}_{n}$ is the $n\times n$ identity matrix; and $\mathcal{CN}(0,\sigma^{2})$ denotes the circularly symmetric complex Gaussian distribution.
}

\section{System Model}

\subsection{System Architecture}

We consider a downlink multi-user \textcolor{black}{MU-}MIMO system with a hybrid beamforming architecture employing movable subarrays. The system comprises a base station (BS) equipped with $C_t$ RF chains, $U$ subarrays, and $M_t$ transmit antennas, with each subarray containing $\frac{M_t}{\textcolor{black}{U}}$ antennas. The BS simultaneously serves $K$ users, each equipped with $N_r$ antennas and receiving $N_s$ data streams. Each subarray forms an $n \times n$ uniform planar array (UPA) with $n=\sqrt{\frac{M_t}{\textcolor{black}{U}}}$.

\textcolor{black}{Each subarray is movable on the $x$-$z$ plane, and its position $\mathbf{p}_u=(x_u,0,z_u)$ is a design variable optimized within a feasible region.} The nominal spacing is $l=\frac{(n-1)\lambda}{2}+l_s$, where $\frac{(n-1)\lambda}{2}$ is the intra-subarray element spacing and $l_s$ is the additional inter-subarray spacing. \textcolor{black}{The initial placement is $\mathbf{p}_1=(0,0,h_t)$ and $x_u=(u_x-1)l,\ z_u=h_t+(u_z-1)l$, where $u_x$ and $u_z$ index the subarray along the $x$- and $z$-axes.}

\subsection{Hybrid Beamforming Architecture}

We adopt the sub-connected architecture where each RF chain connects to one movable subarray such that $C_t=U$.
The received signal at user $k$ is
\begin{equation}
	\begin{aligned}
		{\mathbf{y}}_k = &\mathbf{W}_{BB,k}^H\mathbf{W}_{RF,k}^H\mathbf{H}_k\mathbf{F}_{RF}\mathbf{F}_{BB,k}\mathbf{s}_k \\
		+ &\mathbf{W}_{BB,k}^H\mathbf{W}_{RF,k}^H\mathbf{H}_k\sum_{j\neq k}\mathbf{F}_{RF}\mathbf{F}_{BB,j}\mathbf{s}_j + \mathbf{W}_{BB,k}^H\mathbf{W}_{RF,k}^H\mathbf{n}_k,
	\end{aligned}
\end{equation}
\textcolor{black}{where $\mathbf{W}_{RF,k}$ and $\mathbf{W}_{BB,k}$ denote the analog and digital combiners at user $k$, respectively.}
\textcolor{black}{For analytical simplicity, the receive beamforming is assumed fixed and is absorbed into the effective channel model in the subsequent sum-rate formulation.}
$\mathbf{s}_k \in \mathbb{C}^{\textcolor{black}{N_s} \times 1}$ is the data symbol vector for user $k$,
$\mathbf{F}_{BB} = [\mathbf{F}_{BB,1}, \cdots, \mathbf{F}_{BB,K}] \in \mathbb{C}^{C_t \times \textcolor{black}{K N_s}}$ is the digital precoder,
$\mathbf{F}_{RF} = \mathrm{blkdiag}(\mathbf{F}_{RF,1}, \cdots, \mathbf{F}_{RF,\textcolor{black}{U}}) \in \mathbb{C}^{M_t \times C_t}$ is the block-diagonal analog precoder,
$\mathbf{H}_k \in \mathbb{C}^{\textcolor{black}{N_r} \times M_t}$ is the channel matrix for user $k$, and
$\mathbf{n}_k \sim \mathcal{CN}(\mathbf{0}, \textcolor{black}{\sigma_n^2}\mathbf{I}_{N_r})$ is the additive white Gaussian noise.

\subsection{Cluster-Based Channel Model}

We consider a geometric channel model based on the clustered delay line approach. The channel matrix from the BS to user $k$ is composed of $N_{\text{cl}}$ clusters, each containing $N_{\text{ray}}$ rays:
\begin{equation}
	\begin{aligned}
		\mathbf{H}_k= &\sqrt{\frac{M_t N_r}{N_{\text{cl}} N_{\text{ray}}}} \sum_{i=1}^{N_{\text{cl}}} \sum_{j=1}^{N_{\text{ray}}} \alpha_{i,j}^k  \mathbf{a}_r(\theta_{i,j}^{k,r}, \phi_{i,j}^{k,r}) \\ 
        & \ \ \ \ \ \ \ \ \ \ \ \ \ \ \ \ \ \ \ \ \  \ \times \mathbf{a}_t^H(\theta_{i,j}^t, \phi_{i,j}^t; \mathbf{p}_1, \dots, \mathbf{p}_{\textcolor{black}{U}}),
	\end{aligned}
\end{equation}
where
$\alpha_{i,j}^k$ is the complex gain of the $j$-th ray in the $i$-th cluster for user $k$, $\{\theta_{i,j}^{k,r}, \phi_{i,j}^{k,r}\}$ are the arrival azimuth and elevation angles at user $k$, $\{\theta_{i,j}^t, \phi_{i,j}^t\}$ are the departure azimuth and elevation angles from the BS, $\mathbf{a}_r(\theta, \phi) \in \mathbb{C}^{N_r \times 1}$ is the receive array response vector, and $\mathbf{a}_t(\theta, \phi; \mathbf{p}_1, \dots, \mathbf{p}_{\textcolor{black}{U}}) \in \mathbb{C}^{M_t \times 1}$ is the transmit array response vector.
The transmit array response exhibits a two-level structure due to the subarray configuration:
\begin{equation}
	\mathbf{a}_t(\theta, \phi; \mathbf{p}_1, \dots, \mathbf{p}_{\textcolor{black}{U}}) = \mathbf{a}_{\text{sub}}(\theta, \phi; \mathbf{p}_1, \dots, \mathbf{p}_{\textcolor{black}{U}}) \otimes \mathbf{a}_{\text{elem}}(\theta, \phi),
\end{equation}
where the subarray-level response vector is given by
\begin{equation}
	\mathbf{a}_{\text{sub}}(\theta, \phi; \mathbf{p}_1, \dots, \mathbf{p}_{\textcolor{black}{U}}) = 
	\begin{bmatrix}
		e^{j\frac{2\pi}{\lambda} \mathbf{k}(\theta, \phi)^T \mathbf{m}_1} \\
		e^{j\frac{2\pi}{\lambda} \mathbf{k}(\theta, \phi)^T \mathbf{m}_2} \\
		\vdots \\
		e^{j\frac{2\pi}{\lambda} \mathbf{k}(\theta, \phi)^T \mathbf{m}_{\textcolor{black}{U}}}
	\end{bmatrix}
	\in \mathbb{C}^{\textcolor{black}{U} \times 1},
\end{equation}
with $\mathbf{k}(\theta, \phi) = [\sin\theta\cos\phi, \sin\theta\sin\phi, \cos\theta]^T$ being the wave vector.
The antenna element-level response vector is given by
\begin{equation}	
\mathbf{a}_{\text{elem}}(\theta, \phi) = \mathbf{a}_{\text{elem},x}(\theta, \phi) \otimes \mathbf{a}_{\text{elem},z}(\theta, \phi) \in \mathbb{C}^{\frac{M_t}{\textcolor{black}{U}}\times 1},
\end{equation}
where for a UPA with $n \times n$ elements:
\begin{equation}	
\mathbf{a}_{\text{elem},x}(\theta, \phi) = \frac{1}{\sqrt{n}} \left[1, e^{j\frac{2\pi}{\lambda} d_x \sin\theta\cos\phi}, \dots, e^{j\frac{2\pi}{\lambda} (n-1)d_x \sin\theta\cos\phi} \right]^T,
\end{equation}
\begin{equation}	
\mathbf{a}_{\text{elem},z}(\theta, \phi) = \frac{1}{\sqrt{n}} \left[1, e^{j\frac{2\pi}{\lambda} d_z \cos\theta}, \dots, e^{j\frac{2\pi}{\lambda} (n-1)d_z \cos\theta} \right]^T,
\end{equation}
with $d_x = d_z = \lambda/2$ being the antenna element spacing.

\subsection{Block Diagonalization and Sum-Rate  simplification}

The sum-rate expression is given by 
\begin{equation}
	R_{\text{sum}} = \sum_{k=1}^K \log_2 \left| \textcolor{black}{\mathbf{I}_{N_r}} + \mathbf{R}_k^{-1} \mathbf{H}_k \mathbf{F}_{RF} \mathbf{F}_{BB,k} \mathbf{F}_{BB,k}^H \mathbf{F}_{RF}^H \mathbf{H}_k^H \right|.
\end{equation}
where the interference-plus-noise covariance matrix is:
\begin{equation}
{R}_k = \sum_{j \neq k} \mathbf{H}_k \mathbf{F}_{RF} \mathbf{F}_{BB,j} \mathbf{F}_{BB,j}^H \mathbf{F}_{RF}^H \mathbf{H}_k^H + \sigma^2 \textcolor{black}{\mathbf{I}_{N_r}}.
\end{equation}

For each user $k$, we compute the effective channel as $
	\tilde{\mathbf{H}}_k = \mathbf{H}_k \mathbf{F}_{RF}$.
Then, 
  the interference channel matrix for user $k$ is constructed as $
	\mathbf{H}_{\text{int},k} = [\tilde{\mathbf{H}}_1^T, \dots, \tilde{\mathbf{H}}_{k-1}^T, \tilde{\mathbf{H}}_{k+1}^T, \dots, \tilde{\mathbf{H}}_K^T]^T.$
By Performing singular value decomposition on the interference channel $
	\mathbf{H}_{\text{int},k} = \mathbf{U}_k \mathbf{\Sigma}_k [\mathbf{V}_k^{(1)}, \mathbf{V}_k^{(0)}]^H$,
where $\mathbf{V}_k^{(0)}$ spans the null space of $\mathbf{H}_{\text{int},k}$,
the digital precoder for user $k$ is given by
$
	\mathbf{F}_{BB,k} = \mathbf{V}_k^{(0)} \mathbf{W}_k$,
where $\mathbf{W}_k$ is obtained via SVD of $\tilde{\mathbf{H}}_k \mathbf{V}_k^{(0)}$.
The BD-SVD precoder ensures: $
\mathbf{H}_k \mathbf{F}_{RF} \mathbf{F}_{BB,j} = \mathbf{0},  \forall j \neq k$.
and the sum-rate simplifies to:
\begin{equation}
	\begin{aligned}
		R_{\text{sum}} =& \sum_{k=1}^K \log_2 \left| \mathbf{I}_{N_r} + \frac{1}{\sigma_k^2} \mathbf{H}_k \mathbf{F}_{RF} \mathbf{F}_{BB,k} \mathbf{F}_{BB,k}^H \mathbf{F}_{RF}^H \mathbf{H}_k^H \right| \\
		\overset{(a)}{=}&
		\log_2 \left| \mathbf{I}_{KN_r} + \frac{1}{\sigma_k^2} \mathbf{H} \mathbf{F}_{RF} \mathbf{F}_{BB} \mathbf{F}_{BB}^H \mathbf{F}_{RF}^H \mathbf{H}^H \right|  \\
		=& 	 \log_2 \left| \mathbf{I}_{KN_r} + \frac{1}{\sigma_k^2} \mathbf{H} \mathbf{F}_{RF} \mathbf{F}_{RF}^H \mathbf{H}^H \right|, 
	\end{aligned}
\end{equation}
where $\mathbf{H}=[\mathbf{H}_1^T,\cdots,\mathbf{H}_K^T]^T$, and $(a)$ can be verified via the orthogonality property $
\mathbf{H}_k \mathbf{F}_{RF} \mathbf{F}_{BB,j} = \mathbf{0},  \forall j \neq k$.

\section{Problem Formulation and Solution}

The joint optimization problem for maximizing the sum-rate through subarray positioning and analog precoding is formulated as:
\begin{equation}\label{op}
	\begin{aligned}
		& \underset{\mathbf{x}, \mathbf{z}, \mathbf{F}_{RF}}{\rm arg \ max} 
		& & \log_2 \left| \mathbf{I}_{K N_r} + \frac{1}{\sigma^2} \mathbf{H}(\mathbf{x}, \mathbf{z}) \mathbf{F}_{RF} \mathbf{F}_{RF}^H \mathbf{H}^H(\mathbf{x}, \mathbf{z}) \right| \\
		& \text{subject to}
		& & \sqrt{(x_u - x_v)^2 + (z_u - z_v)^2} \geq D_{\text{min}}, \quad \forall u \neq v, \\
		&&& \mathbf{F}_{RF} = \mathrm{blkdiag}(\mathbf{F}_{RF,1}, \dots, \mathbf{F}_{RF,U}), \\
        &&& \textcolor{black}{|[\mathbf{F}_{RF,u}]_{i,j}| = 1, \quad \forall u,\ \forall i,j,} \\
		&&& \|\mathbf{F}_{RF}\mathbf{F}_{BB}\|_F^2 \leq P_{\text{max}},
	\end{aligned}
\end{equation}
where \textcolor{black}{$\mathbf{x} = [x_1, \dots, x_U]^T$ and $\mathbf{z} = [z_1, \dots, z_U]^T$} represent the position vectors of subarrays, $D_{\text{min}} $ is the minimum spacing between subarrays, and $P_{\text{max}}$ is the maximum transmit power constraint.

To solve the sum-rate maximization problem, we employ an alternating optimization method for antenna coefficients and subarray positions under the SIC framework.
\textcolor{black}{For practical implementation, each subarray position is constrained within a feasible movable region (e.g., $x_u \leq x_{\max}$) and can be selected from a set of predefined sampling points.}

\subsection{Analog \textcolor{black}{B}eamforming \textcolor{black}{O}ptimization}

The joint optimization of all RF chains under the non-convex constant modulus constraint is intractable.
We adopt a successive optimization approach, designing one RF chain at a time while treating the signals from the others as interference.
\textcolor{black}{We follow a cyclic SIC/BCD update: when optimizing $\mathbf{F}_{RF,m}$, the remaining $\{\mathbf{F}_{RF,i}\}_{i\neq m}$ are fixed to their latest values (initialized by a feasible phase-only design) and updated sequentially for $m=1,\ldots,U$ until convergence.}

We decompose the total precoding covariance matrix as
\begin{equation}
	\mathbf{F}_{RF} \mathbf{F}_{RF}^H = \mathbf{F}_{RF,m} \mathbf{F}_{RF,m}^H
    + \sum_{i \neq m}^{\textcolor{black}{U}} \mathbf{F}_{RF,i} \mathbf{F}_{RF,i}^H,
\end{equation}
where $\mathbf{F}_{RF,m}$ is the precoding vector of the $m$-th RF chain. Substituting into (1), the spectral efficiency can be rewritten as
\begin{equation}
	\begin{aligned}
		R = \log_2 \left| \mathbf{I}_{KN_r} + \mathbf{B}_m + \mathbf{A}_m \right|,
	\end{aligned}
\end{equation}
where
\begin{equation}
	\mathbf{A}_m = \frac{1}{\sigma^2} \mathbf{H} \left( \sum_{i \neq m}^{\textcolor{black}{U}} \mathbf{F}_{RF,i} \mathbf{F}_{RF,i}^H \right) \mathbf{H}^H,
\end{equation}
and the desired signal covariance matrix is denoted by
\begin{equation}
\end{equation}

Applying the matrix determinant lemma, $\left| \mathbf{I} + \mathbf{A} + \mathbf{B} \right| = \left| \mathbf{I} + \mathbf{A} \right| \cdot \left| \mathbf{I} + (\mathbf{I} + \mathbf{A})^{-1} \mathbf{B} \right|$, the rate expression is decomposed as:
\begin{equation}
	R = \underbrace{\log_2 \left| \mathbf{I}_{KN_r} + \mathbf{A}_m \right|}_{R_1} 
    + \underbrace{\log_2 \left| \mathbf{I}_{KN_r} + (\mathbf{I}_{KN_r} + \mathbf{A}_m)^{-1} \mathbf{B}_m \right|}_{R_2}.
\end{equation}

The term $R_1$ is independent of $\mathbf{F}_{RF,m}$. Therefore, maximizing $R$ with respect to $\mathbf{F}_{RF,m}$ is equivalent to maximizing $R_2$.
\textcolor{black}{This successive update is not globally optimal due to the coupled constant-modulus constraints, but it provides a low-complexity stationary solution.}

Applying the identity $\left| \mathbf{I} + \mathbf{X}\mathbf{X}^H \right| = \left| \mathbf{I} + \mathbf{X}^H\mathbf{X} \right|$, we can reformulate $R_2$ as
\begin{equation}
\begin{aligned}
	R_2 &= \log_2 \left| \mathbf{I}_{KN_r} + (\mathbf{I}_{KN_r} + \mathbf{A}_m)^{-1/2} \mathbf{B}_m (\mathbf{I}_{KN_r} + \mathbf{A}_m)^{-1/2} \right| \\
	&= \log_2 \left| \textcolor{black}{\mathbf{I}_{N_s}} + \frac{1}{\sigma^2} \mathbf{F}_{RF,m}^H 
    \underbrace{\mathbf{H}^H (\mathbf{I}_{KN_r} + \mathbf{A}_m)^{-1} \mathbf{H}}_{\mathbf{C}_m} 
    \mathbf{F}_{RF,m} \right|,
\end{aligned}
\end{equation}
where we have defined the effective channel matrix:
\begin{equation}
	\mathbf{C}_m = \mathbf{H}^H (\mathbf{I}_{KN_r} + \mathbf{A}_m)^{-1} \mathbf{H}.
\end{equation}

Ignoring the constant modulus constraint, the unconstrained solution that maximizes (5) is found by performing the SVD of $\mathbf{C}_m$:
\begin{equation}
	\mathbf{C}_m = \mathbf{U}_m \boldsymbol{\Sigma}_m \mathbf{U}_m^H,
\end{equation}
where $\mathbf{U}_m$ is a unitary matrix and $\boldsymbol{\Sigma}_m$ is a diagonal matrix of singular values in descending order. 
The optimal unconstrained precoder for the $m$-th chain is then given by the first $N_s$ dominant singular vectors:
$ \tilde{\mathbf{F}}_{RF,m}^\star = \mathbf{U}_m(:, 1:N_s)$.

The solution above does not satisfy the constant modulus constraint. The standard approach is to project the unconstrained solution onto the feasible set of phase shifters:
\begin{equation}
\mathbf{F}_{RF,m} = e^{ j  \angle\left( \mathbf{U}_m(:, 1:N_s) \right) } .
\end{equation}
\textcolor{black}{Since $|e^{j\theta}|=1$, this projection ensures $|[\mathbf{F}_{RF,u}]_{i,j}|=1$ for each block $\mathbf{F}_{RF,u}$ under the sub-connected $\mathrm{blkdiag}(\cdot)$ architecture.}

\subsection{Subarray Position Optimization}

We re-organize the channel matrix and analog beamforming matrix as $
	\mathbf{H}= [\mathbf{H}_1,\cdots,\mathbf{H}_m,\cdots,\mathbf{H}_M]$,
and 
\begin{equation}
	\mathbf{F}_{RF} =\begin{bmatrix}
		 \overline{\mathbf{F}}_{RF,1}\\ \vdots \\ \overline{\mathbf{F}}_{RF,M}
	\end{bmatrix}.
\end{equation}

Noticing $\overline{\mathbf{F}}_{RF,m}\overline{\mathbf{F}}_{RF,i}^H=\mathbf{0}$ for $i\neq m$, we obtain
\begin{equation}
\begin{aligned}
&\mathbf{H}\mathbf{F}_{RF}\mathbf{F}_{RF}^H\mathbf{H}^H=\\ 
&\left(\mathbf{H}_m \overline{\mathbf{F}}_{RF,m} + \sum_{i\neq m}\mathbf{H}_{i} \overline{\mathbf{F}}_{RF,i}\right)\left(\mathbf{H}_m \overline{\mathbf{F}}_{RF,m} + \sum_{i\neq m}\mathbf{H}_{i} \overline{\mathbf{F}}_{RF,i}\right)^H\\
=&\mathbf{H}_m \overline{\mathbf{F}}_{RF,m}\overline{\mathbf{F}}_{RF,m}^H\mathbf{H}_m^H+\sum_{i\neq m}^{M}\mathbf{H}_i \overline{\mathbf{F}}_{RF,i}\overline{\mathbf{F}}_{RF,i}^H\mathbf{H}_i^H. 
\end{aligned}
\end{equation}
 
Hence, the spectral efficiency can be written as
\begin{equation}
	\begin{aligned}
		&
		\log_2 \left| \mathbf{I}_{KN_r} + \frac{1}{\sigma_k^2} \mathbf{H} \mathbf{F}_{RF} \mathbf{F}_{RF}^H \mathbf{H}^H \right|
        = \log_2 \left| \mathbf{I}_{KN_r} + \mathbf{A}_m^\prime \right| \\
		&\ \ \ \ \ \  \ \ \ \   \ \ \     
        + \log_2 \left|  \mathbf{I}_{KN_r} +   (\mathbf{I}+\mathbf{A}_m^\prime) \mathbf{H}_m \overline{\mathbf{F}}_{RF,m}\overline{\mathbf{F}}_{RF,m}^H\mathbf{H}_m^H
		\right|,
	\end{aligned}
\end{equation}
where $\mathbf{A}_m^\prime=\sum_{i\neq m}^{M}\mathbf{H}_i \overline{\mathbf{F}}_{RF,i}\overline{\mathbf{F}}_{RF,i}^H\mathbf{H}_i^H$  

According to the SIC procedure, all subarrays' positions can be optimized alternatively.
Therefore, the optimization problem with respect to the $m$th subarray's position is given by
\begin{equation}
	\begin{aligned}
		\underset{{x}_m,{z}_m} {\rm arg \ max} \ &
		\log_2 \left|  \mathbf{I}_{KN_r} +   (\mathbf{I}+\mathbf{A}_m^\prime) \mathbf{H}_m \mathbf{F}_{RF,m}\mathbf{F}_{RF,m}^H\mathbf{H}_m^H
		\right|, \\
		&
		{\rm s.t.} \ \sqrt{(x_m-x_i)^2+(z_m-z_i)^2}\geq D_{\rm min}, \forall  i\neq m.
	\end{aligned}
\end{equation}

\textcolor{black}{Although inter-subarray spacing and movable-region constraints exist, we address them by design: partition the movable area into $M$ disjoint subregions and confine each subarray to its own region. This guarantees the minimum inter-subarray distance without pairwise constraints, so each update is an unconstrained local search and Nelder--Mead is applicable.}

The algorithm begins by constructing an initial simplex \( S^{(0)} = \{\mathbf{p}_1, \mathbf{p}_2, \mathbf{p}_3\} \), where each vertex \( \mathbf{p}_i = [x_i, z_i]^T \) represents a coordinate pair.
\textcolor{black}{The simplex is initialized as an isosceles right triangle (or equilateral triangle) with unit side length.}
The vertices are ordered such that \( f(\mathbf{p}_1) \leq f(\mathbf{p}_2) \leq f(\mathbf{p}_3) \), meaning \( \mathbf{p}_3 \) is the current best point.

The algorithm then iteratively refines the simplex. The centroid \( \overline{\mathbf{p}} = 0.5(\mathbf{p}_2 + \mathbf{p}_3) \) is computed, and the worst vertex \( \mathbf{p}_1 \) is reflected to generate \( \mathbf{p}_r = \overline{\mathbf{p}} + \alpha (\overline{\mathbf{p}} - \mathbf{p}_1) \) with \( \alpha = 1 \).
\textcolor{black}{If \( f(\mathbf{p}_r) > f(\mathbf{p}_3) \), perform expansion; if \( f(\mathbf{p}_r) \leq f(\mathbf{p}_2) \), perform contraction; otherwise accept reflection.}
If contraction fails, shrink towards \( \mathbf{p}_3 \) via \( \mathbf{p}_i' = \mathbf{p}_3 + \delta (\mathbf{p}_i - \mathbf{p}_3) \) for \( i=1,2 \), with \( \delta = 0.5 \).
\textcolor{black}{All trial points are kept within the assigned subregion (e.g., by simple clipping/projection) before evaluating $f(\cdot)$.}
In this way, the simplex operations proceed normally while feasibility is preserved throughout the iterations.

\section{Simulation Results}
This section \textcolor{black}{evaluates the proposed methods via simulations} at a carrier frequency of $28$~GHz. 
We consider \textcolor{black}{$K=4$ users and $U=4$ subarrays, each user receiving $N_s=2$ streams, with $M_t=64$ transmit antennas.} 
The channel uses a clustered model with $N_{cl}\in\{2,8\}$ and $N_{ray}\in\{5,10\}$; cluster angles are uniformly distributed with an angular spread of $15^\circ$, and path gains follow $\mathcal{CN}(0,1)$. 
The evaluated schemes include: 
\begin{itemize}
\item 	\textbf{SIC-FPA}: This method adopts the BD approach for digital beamforming and the SIC approach for analog beamforming under the modulus-1 beamforming constraint.
\item 	\textbf{SIC-MA}: This method adopts the BD approach for digital beamforming, and the SIC approach for subarray position optimization and analog beamforming under the modulus-1 beamforming constraint.
\item 	\textbf{Unconstrained SIC-FPA}: This method adopts the BD approach for digital beamforming and the SIC approach for analog beamforming.
\item 	\textbf{Unconstrained SIC-MA}: This method adopts the BD approach for digital beamforming, and the SIC approach for subarray position optimization and analog beamforming.
\end{itemize}

\textcolor{black}{The proposed SIC-based alternating optimization has per-iteration complexity on the order of $O(K^2U)$ (up to standard matrix operations), which is significantly lower than fully joint optimization.}

\begin{figure}
	\centering 
	\includegraphics[width=3.25in]{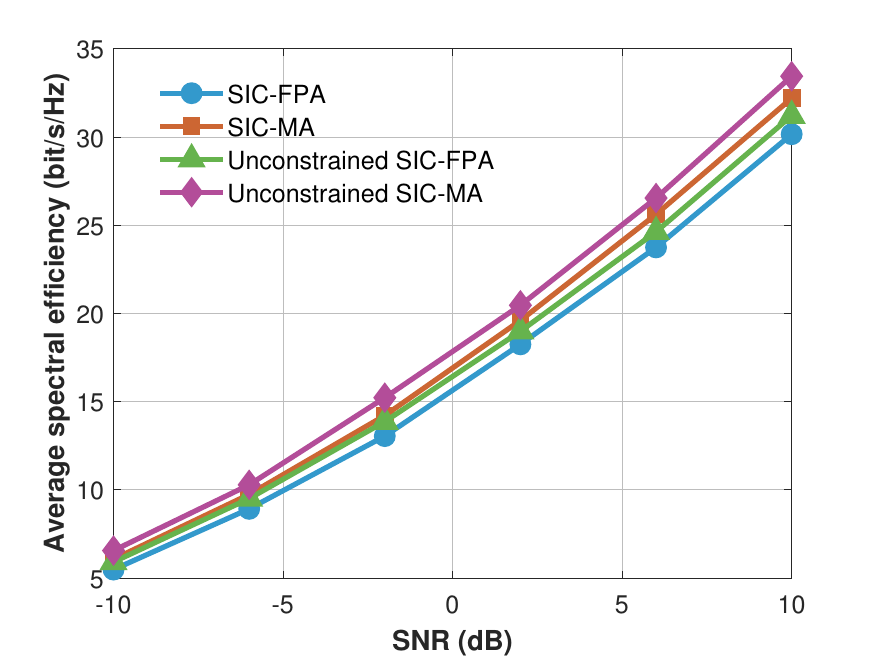}
	\caption{The rate of different cases versus SNR.}\label{SNR} 
\end{figure} 

Fig. \ref{SNR} evaluates the impact of SNR on different schemes in terms of the spectral efficiency, where $N_{\rm cl}=2$, $N_{\rm ray}=5$, and the movable region is defined by a square region in which the length is set to $12\lambda$. 
It can be observed that MA-based schemes outperform FPA-based schemes. For instance, SIC-MA shows a performance improvement at any SNR levels.

\begin{figure}
	\centering 
	\includegraphics[width=3.25in]{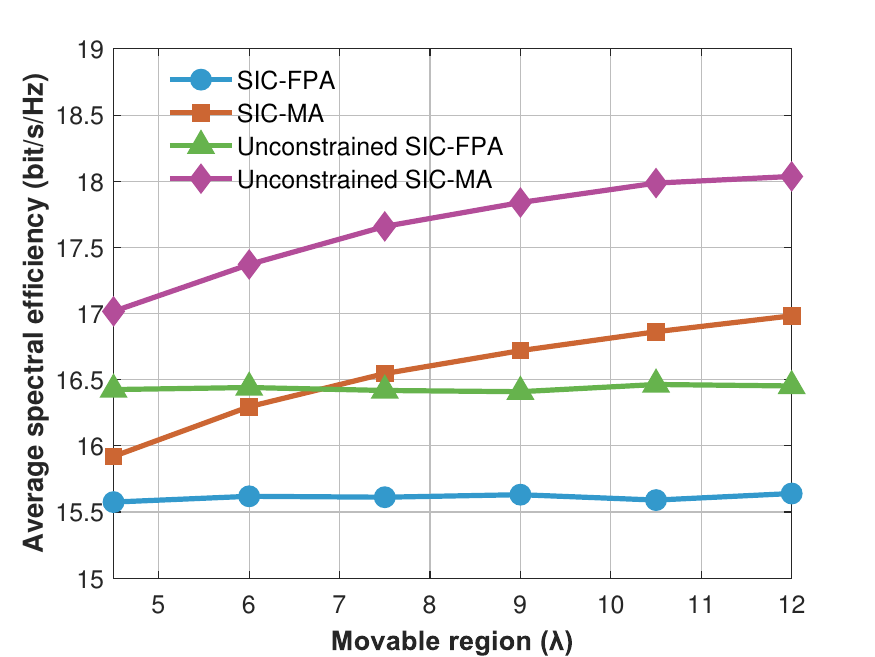}
	\caption{Spectral efficiency varying with the movable region, $N_{\rm cl}=2$, $N_{\rm ray}=5$.}\label{M1} 
\end{figure} 

In Fig.~\ref{M1}, we set $\mathrm{SNR}=0$~dB, the number of channel clusters and rays are set to $N_{\rm cl}=2$ and $N_{\rm ray}=5$, and the movable region is defined by a square region in which the length \textcolor{black}{varies} from $4.5\lambda$ to $12\lambda$. 
As the movable region enlarges, MA-based schemes benefit from increased positional freedom, yielding higher array gain and improved interference management. 
In particular, SIC-MA consistently outperforms SIC-FPA, confirming the advantage granted by the flexibility in sub-array placement, thereby improving the effective channel conditions and \textcolor{black}{SIC interference cancellation}.

\begin{figure}
	\centering 
	\includegraphics[width=3.25in]{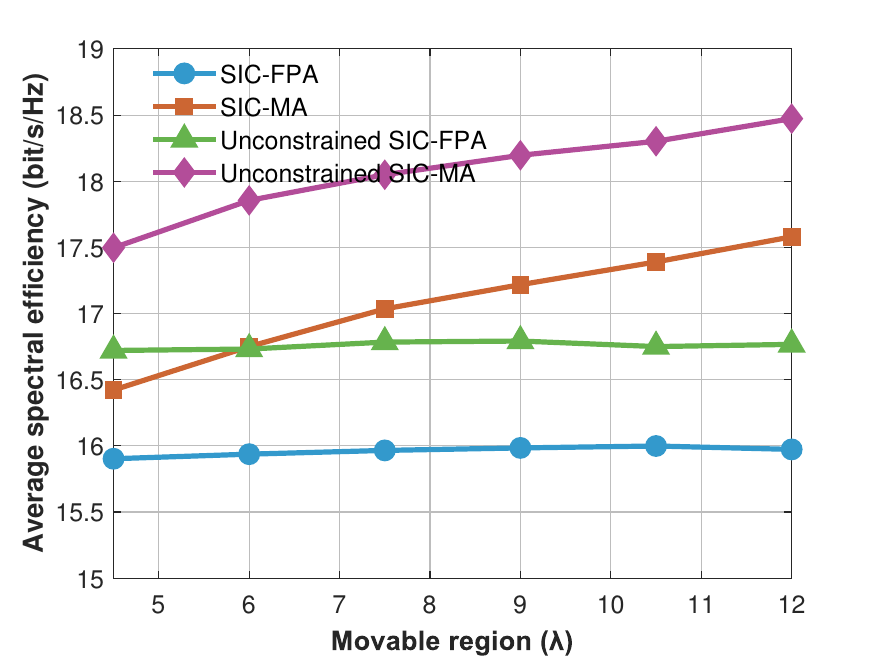}
	\caption{Spectral efficiency varying with the movable region, $N_{\rm cl}=8$, $N_{\rm ray}=10$.}\label{M2} 
\end{figure}

The unconstrained SIC-MA provides an upper bound, and the gap to SIC-MA quantifies the loss due to practical constraints; a similar upper-bound relation holds for SIC-FPA versus unconstrained SIC-FPA. 
Fig.~\ref{M2} shows the same trend as Fig.~\ref{M1}, and a larger number of paths further improves the spectral efficiency.

\section{Conclusion}

In this study, we have investigated the integration of subarray-level movable antennas into an MU-MIMO system to unlock the potential of spatial degrees of freedom beyond traditional beamforming. The core of our work was the joint optimization of hybrid beamforming and subarray positions to maximize spectral efficiency, a complex problem decomposed via a BD-based precoding framework and solved with a dedicated SIC-based algorithm for the analog and positional components.
The simulation results provide conclusive evidence for the superiority of the proposed SIC-MA architecture. It consistently achieves higher spectral efficiency compared to the fixed subarray position (SIC-FPA) benchmark across a wide range of movable regions. The observed monotonic performance improvement with an increasing movable region establishes a clear trade-off between the system's physical configuration flexibility and its communication performance.
\textcolor{black}{We note that subarray repositioning may introduce additional actuation energy and hardware/control cost, which is an important practical consideration for future work.}
Future research may explore more sophisticated beamforming schemes, the impact of channel estimation errors on position optimization, and the development of low-complexity algorithms for real-time deployment in dynamic wireless environments.

\vspace{12pt}
\bibliographystyle{IEEEtran}

\end{document}